\documentstyle[prd,aps,amssymb,preprint,epsfig]{revtex}
\input prepictex
\input pictex
\input postpictex

\begin{document}


\def\calA{{\cal A}}
\def\calB{{\cal B}}
\def\calH{{\cal H}}
\def\calO{{\cal O}}

\def\cbar{{\bar c}}
\def\nbar{{\bar n}}
\def\qbar{{\bar q}}
\def\sbar{{\bar s}}
\def\ubar{{\bar u}}


\def\etal{{\it et al.}}
\def\ibid#1#2#3{{\it ibid.} {\bf #1}, #3 (#2)}

\def\epjc#1#2#3{Eur. Phys. J. C {\bf #1}, #3 (#2)}
\def\ijmpa#1#2#3{Int. J. Mod. Phys. A {\bf #1}, #3 (#2)}
\def\jhep#1#2#3{J. High Energy Phys. {\bf #1}, #3 (#2)}
\def\mpl#1#2#3{Mod. Phys. Lett. A {\bf #1}, #3 (#2)}
\def\npb#1#2#3{Nucl. Phys. {\bf B#1}, #3 (#2)}
\def\plb#1#2#3{Phys. Lett. B {\bf #1}, #3 (#2)}
\def\prd#1#2#3{Phys. Rev. D {\bf #1}, #3 (#2)}
\def\prl#1#2#3{Phys. Rev. Lett. {\bf #1}, #3 (#2)}
\def\rep#1#2#3{Phys. Rep. {\bf #1}, #3 (#2)}
\def\zpc#1#2#3{Z. Phys. {\bf #1}, #3 (#2)}


\title{Radiative $B$ decays to the axial $K$ mesons at next-to-leading order}
\author{Jong-Phil Lee\footnote{e-mail: jplee@phya.yonsei.ac.kr}}
\address{Department of Physics and IPAP, Yonsei University, Seoul, 120-749, Korea}

\tighten
\maketitle

\begin{abstract}
We calculate the branching ratios of $B\to K_1\gamma$ at next-to-leading order
(NLO) of $\alpha_s$ where $K_1$ is the orbitally excited axial vector meson.
The NLO decay amplitude is divided into the vertex correction and the hard
spectator interaction part.
The one is proportional to the weak form factor of $B\to K_1$ transition while
the other is a convolution between light-cone distribution amplitudes and hard
scattering kernel.
Using the light-cone sum rule results for the form factor, we have
$\calB(B^0\to K_1^0(1270)\gamma)=(0.828\pm0.335)\times 10^{-5}$ and
$\calB(B^0\to K_1^0(1400)\gamma)=(0.393\pm0.151)\times 10^{-5}$.
\end{abstract}
\pagebreak

\section{Introduction}

Radiative $B$ decays into Kaons provide abundant issues for both theorists and
experimentalists. 
After the first measurement at CLEO, $B\to K^*\gamma$ is now also measured in
Belle and BaBar:
\begin{eqnarray}
\calB(B^0\to K^{*0}\gamma)&=&\left\{\begin{array}{cc}
 (4.09\pm0.21\pm0.19)\times 10^{-5} & {\rm Belle}~\cite{Belle} \\
 (4.23\pm0.40\pm0.22)\times 10^{-5} & {\rm BaBar}~\cite{BaBar} \\
 (4.55\pm0.70\pm0.34)\times 10^{-5} & {\rm CLEO}~\cite{CLEO}\end{array}\right.~,
\\
\calB(B^+\to K^{*+}\gamma)&=&\left\{\begin{array}{cc}
 (4.40\pm0.33\pm0.24)\times 10^{-5} & {\rm Belle}~\cite{Belle} \\
 (3.83\pm0.62\pm0.22)\times 10^{-5} & {\rm BaBar}~\cite{BaBar} \\
 (3.76\pm0.86\pm0.28)\times 10^{-5} & {\rm CLEO}~\cite{CLEO}\end{array}\right.~.
\end{eqnarray}
Theoretical advances in $B\to K^*\gamma$ have been noticeable for a decade.
QCD corrections at next-to-leading order (NLO) of $\calO(\alpha_s)$ was already
considered in \cite{Soares,Greub,Hurth}.
Furthermore, relevant Wilson coefficients have been improved 
\cite{Adel,Chetyrkin} up to three-loop calculations.
Recent developments of the QCD factorization \cite{BBNS} helped one calculate
the hard spectator contributions systematically in a factorized form through
the convolution at the heavy quark limit \cite{Feldmann,Siedel,Bosch}.
$B\to K^*\gamma$ is also analyzed in the effective theories at NLO, such as
large energy effective theory \cite{Ali} and the soft-collinear effective
theory (SCET) \cite{Chay}.
\par
In addition to $K^*$, higher resonances of Kaon also deserve much attention.
Especially, it was suggested that $B\to K_{\rm res}(\to K\pi\pi)\gamma$ can
provide a direct measurement of the photon polarization \cite{Gronau}.
In particular, it was shown that $B\to K_1(1400)\gamma$ can produce large 
polarization asymmetry of $\approx 33\%$ in the standard model.
In the presence of anomalous right-handed couplings, the polarization can
be severely reduced in the parameter space allowed by current experimental
bounds of $B\to X_s\gamma$ \cite{jplee}.
It was also argued that the $B$ factories can now make a lot of $B{\bar B}$
pairs enough to check the anomalous couplings through the measurement of the
photon polarization.
\par
As for the axial $K_1$,
unfortunately, current measurements give only upper bounds for 
$B\to K_1\gamma$ \cite{Belle2}:
\begin{eqnarray}
\calB(B^+\to K_1^+(1270)\gamma)&<&9.9\times 10^{-5}~,\\
\calB(B^+\to K_1^+(1400)\gamma)&<&5.0\times 10^{-5}~.
\end{eqnarray}
For the decays of $B\to K_2(1430)\gamma$, CLEO and the $B$ factories have 
reported the branching ratios
\begin{eqnarray}
\calB(B\to K_2^*\gamma)&=&(1.66^{+0.59}_{-0.53}\pm0.13)\times 10^{-5}~
{\rm CLEO}~\cite{CLEO}~,\\
\calB(B^0\to K_2^{*0}\gamma)&=&\left\{\begin{array}{cc}
(1.3\pm0.5\pm0.1)\times 10^{-5} & {\rm Belle}~\cite{Belle2}\\
(1.22\pm0.25\pm0.11)\times 10^{-5} & {\rm BaBar}~\cite{BaBar2}
\end{array}\right.~,\\
\calB(B^+\to K_2^{*+}\gamma)&=&(1.44\pm0.40\pm0.13)\times 10^{-5}~
{\rm BaBar}~\cite{BaBar2}~.
\end{eqnarray}
Since the higher resonant Kaons are rather heavy $\gtrsim 1~{\rm GeV}$, it is
quite natural and attractive to consider them as heavy mesons.
The advent of heavy quark effective theory (HQET) provoked many studies.
Although the HQET simplifies the analysis by reducing number of the 
independent form factors involved, other non-perturbative methods are needed
to complete the phenomenological explanation.
These HQET-based analyses include HQET-ISGW (Isgur-Scora-Grinstein-Wise) 
\cite{Ohl} and HQET-NRQM (Non-Relativistic Quark Model) \cite{Veseli}.
Other model calculations have been done in \cite{Altomari,Atwood,Ebert,HYCheng}.
\par
In this paper, the branching ratios of $B\to K_1\gamma$ at NLO of $\alpha_s$
are calculated.
We adopt the QCD factorization framework where the hard spectator interactions 
are described by the convolution between the hard-scattering kernel and the 
lint-cone distribution amplitudes (DA) at the heavy quark limit.
All the non-perturbative nature are encapsulated in the DA
while the hard kernel is perturbatively calculable.
Basically, $B\to K_1\gamma$ shares many things with $B\to K^*\gamma$.
Only the difference is the DA for the daughter mesons.
Vector and axial vector mesons are distinguished by the $\gamma_5$ in the gamma
structure of DA and some non-perturbative parameters.
But the presence of $\gamma_5$ does not alter the calculation, giving the same 
result for the perturbative part.
As for the non-perturbative parameters, the decay constant is most important.
If higher twist terms are included, the Gegenbauer moments in the Gegenbauer
expansion are also process dependent.
We will not consider higher twists for simplicity.
\par
Another NLO contributions are the vertex corrections to the relevant operators.
They are all proportional to the leading operator $O_7$.
The matrix elements of $O_7$ are parameterized by several form factors.
For the radiative decays where the emitted photons are real, only one form
factor enters the decay amplitude.
However,
other non-perturbative calculation is needed for the value of the form factor.
We use the light-cone sum rule (LCSR) results for it \cite{Safir}.
\par
Thus at NLO, $B\to K^*\gamma$ and $B\to K_1\gamma$ are characterized by the
weak form factor $F^{V(A)}_+$ and decay constant, plugged by the common
perturbative and kinematical factors.
With $\calB(B\to K^*\gamma)$ at hand, near future measurements of 
$B\to K_1\gamma$ will check this structure.
\par
The paper is organized as follows.
General setup and leading contribution to $B\to K_1\gamma$ are given in the
next Section.
Section III is devoted to the NLO corrections.
The resulting branching ratios and related discussions appear in Sec.\ IV.
We conclude in Sec.\ V.

\section{Leading order contribution}

Let us start with the effective Hamiltonian for $b\to s\gamma$,
\begin{equation}
\calH_{\rm eff}(b\to s\gamma)=-\frac{G_F}{\sqrt{2}}V_{tb}V_{ts}^*
 \sum_{i=1}^{8}C_i(\mu)O_i(\mu)~,
\end{equation}
where
\begin{eqnarray}
O_1&=&(\sbar_i c_j)_{V-A}(\cbar_j b_i)_{V-A}~, \nonumber\\
O_2&=&(\sbar_i c_i)_{V-A}(\cbar_j b_j)_{V-A}~, \nonumber\\
O_3&=&(\sbar_i b_i)_{V-A}\sum_q (\qbar_j q_j)_{V-A}~, \nonumber\\
O_4&=&(\sbar_i b_j)_{V-A}\sum_q (\qbar_j q_i)_{V-A}~, \nonumber\\
O_5&=&(\sbar_i b_i)_{V-A}\sum_q (\qbar_j q_j)_{V+A}~, \nonumber\\
O_6&=&(\sbar_i b_j)_{V-A}\sum_q (\qbar_j q_i)_{V+A}~, \nonumber\\
O_7&=&\frac{em_b}{8\pi^2}\sbar_i\sigma^{\mu\nu}(1+\gamma_5)b_i F_{\mu\nu}~, \nonumber\\
O_8&=&\frac{g_sm_b}{8\pi^2}\sbar_i\sigma^{\mu\nu}(1+\gamma_5)T^a_{ij}b_j G^a_{\mu\nu}~. 
\end{eqnarray}
Here $i,j$ are color indices, and we neglect the CKM element $V_{ub}V_{us}^*$ 
as well as the $s$-quark mass.
The leading contribution to $B\to K_1\gamma$ comes from the electromagnetic
operator $O_7$ as shown in Fig.\ 1.
The matrix element of $O_7$ is described by the transition form factors
$F^A_{\pm,0}$ which are defined by
\begin{mathletters}
\begin{eqnarray}
\lefteqn{
\langle K_1(p',\epsilon)|\sbar i\sigma_{\mu\nu}q^\nu b|B(p)\rangle}\nonumber\\
&=&
F^A_+(q^2)\Big[(\epsilon^*\cdot q)(p+p')_\mu-\epsilon^*_\mu(p^2-p^{\prime 2})\Big]
+F^A_-(q^2)\Big[(\epsilon^*\cdot q)q_\mu-\epsilon^*_\mu q^2\Big]\nonumber\\
&&+\frac{F^A_0(q^2)\epsilon^*\cdot q}{m_B m}\Big[
(p^2-p^{\prime 2})q_\mu-(p+p')_\mu q^2\Big]~,\\
\lefteqn{
\langle K_1(p',\epsilon)|\sbar i\sigma_{\mu\nu}\gamma_5q^\nu b|B(p)\rangle=
iF^A_+(q^2)\epsilon_{\mu\nu\alpha\beta}\epsilon^{*\nu}q^\alpha(p+p')^\beta~,}
\end{eqnarray}
\end{mathletters}
where $m$ and $\epsilon^\mu$ are the mass and polarization vector of $K_1$,
respectively, and $q=p-p'$ is the photon momentum.
In case of real photon emission ($q^2=0$), only $F^A_+$ is involved as
\begin{eqnarray}
\langle O_7\rangle_A &\equiv&
\langle K_1(p',\epsilon)\gamma(q,e)|O_7|B(p)\rangle\nonumber\\
&=&\frac{em_b}{4\pi^2}F^A_+(0)\Big[\epsilon^*\cdot q (p+p')\cdot e^*
   -\epsilon^*\cdot e^* (p^2-p^{\prime 2})
   +i\epsilon_{\mu\nu\alpha\beta}e^{*\mu}\epsilon^{*\nu}q^\alpha(p+p')^\beta
 \Big]~,
\end{eqnarray}
with $e^\mu$ being the photon polarization vector.
The decay rate is straightforwardly obtained to be
\begin{equation}
\Gamma(B\to K_1\gamma)=\frac{G_F^2\alpha m_b^2 m_B^3}{32\pi^4}|V_{tb}V_{ts}^*|^2
 \Bigg(1-\frac{m^2}{m_B^2}\Bigg)^3|F^A_+|^2 |C_7^{\rm eff (0)}|^2~,
\end{equation}
where $\alpha$ is the fine-structure constant and $C_7^{\rm eff(0)}$ is the
effective Wilson coefficient at leading order.

\section{Matrix elements at next-to-leading order of $\calO(\alpha_s)$}

\begin{figure}
\vskip 2cm
\begin{center}
\epsfig{file=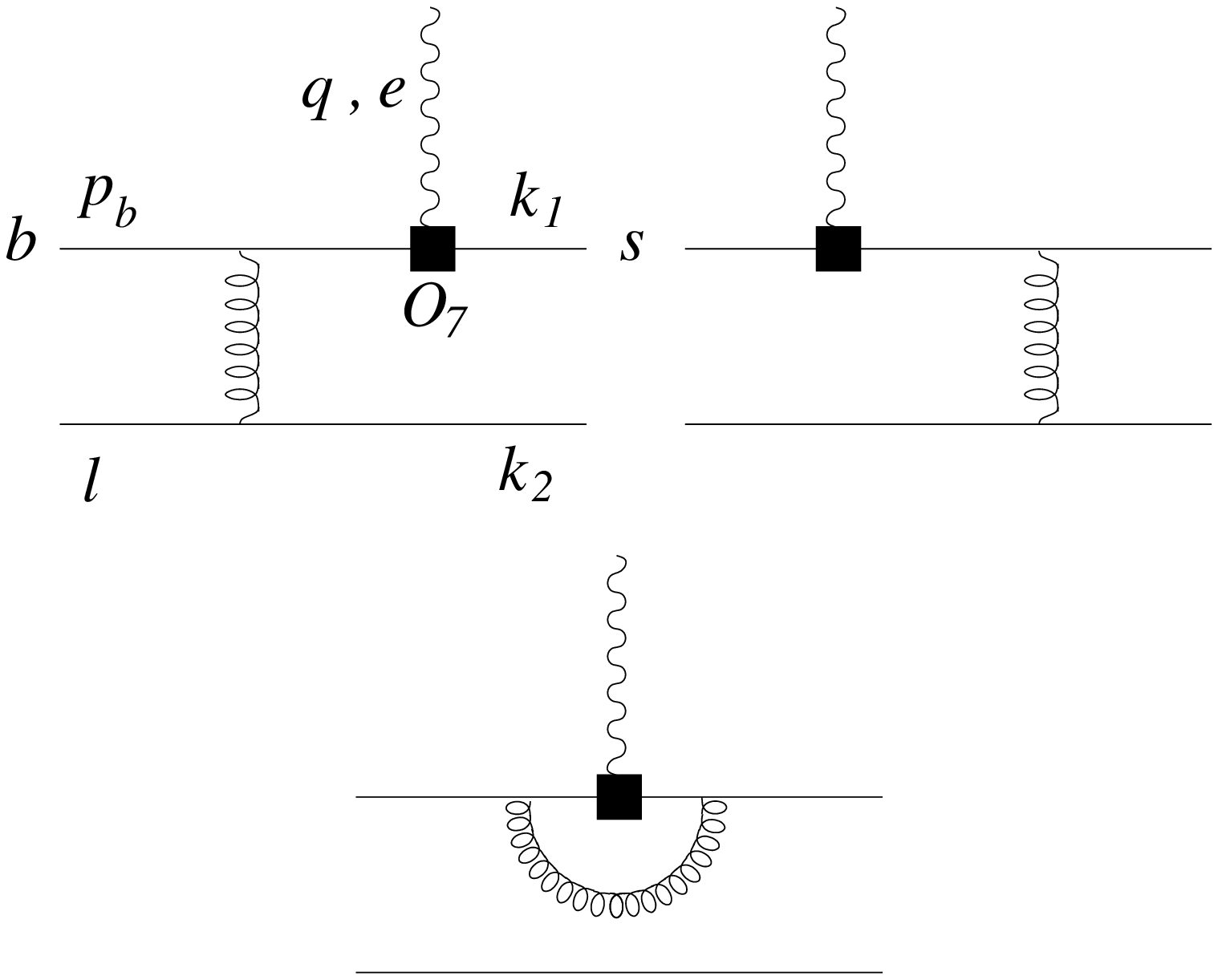}
\end{center}
\caption{}
\label{O7NLO}
\end{figure}

At next-to-leading order of $\alpha_s$, there are other contributions from
the operators $O_2$ and $O_8$. 
We simply neglect the annihilation topologies.
Explicitly, the decay amplitude $\calA$ is given by
\begin{equation}
\calA(B\to K_1\gamma)=-\frac{G_F}{\sqrt{2}}V_{tb}V_{ts}^*(
 C_7^{\rm eff}\langle O_7\rangle+C_2\langle O_2\rangle
 +C_8^{\rm eff}\langle O_8\rangle)~,
\end{equation}
where $\langle O_i\rangle\equiv \langle K_1\gamma|O_i|B\rangle$.
Every $\langle O_i\rangle$ has its vertex correction 
$\langle O_i\rangle_{VC}$ and hard spectator interaction term 
$\langle O_i\rangle_{HS}$ as shown in Figs.\ 2 and 3;
\begin{equation}
\langle O_i\rangle=\langle O_i\rangle_{VC}+\langle O_i\rangle_{HS}~.
\end{equation}
As for $\langle O_7\rangle$, all the subleading contributions shown in Fig.\ 
\ref{O7NLO} are absorbed into
the form factor $F^A_+$ while the corresponding Wilson coefficient 
$C_7^{\rm eff}$ contains its NLO part,
\begin{equation}
C_7^{\rm eff}(\mu)=C_7^{\rm eff(0)}(\mu)
 +\frac{\alpha_s(\mu)}{4\pi}C_7^{\rm eff(1)}(\mu)~.
\end{equation}
On the other hand, the leading order $C_2^{(0)}$ and $C_8^{\rm eff(0)}$
are sufficient for $C_2$ and $C_8$ since $O_2$ and $O_8$ contributions begin
at NLO.
\par
\begin{figure}
\begin{center}
\epsfig{file=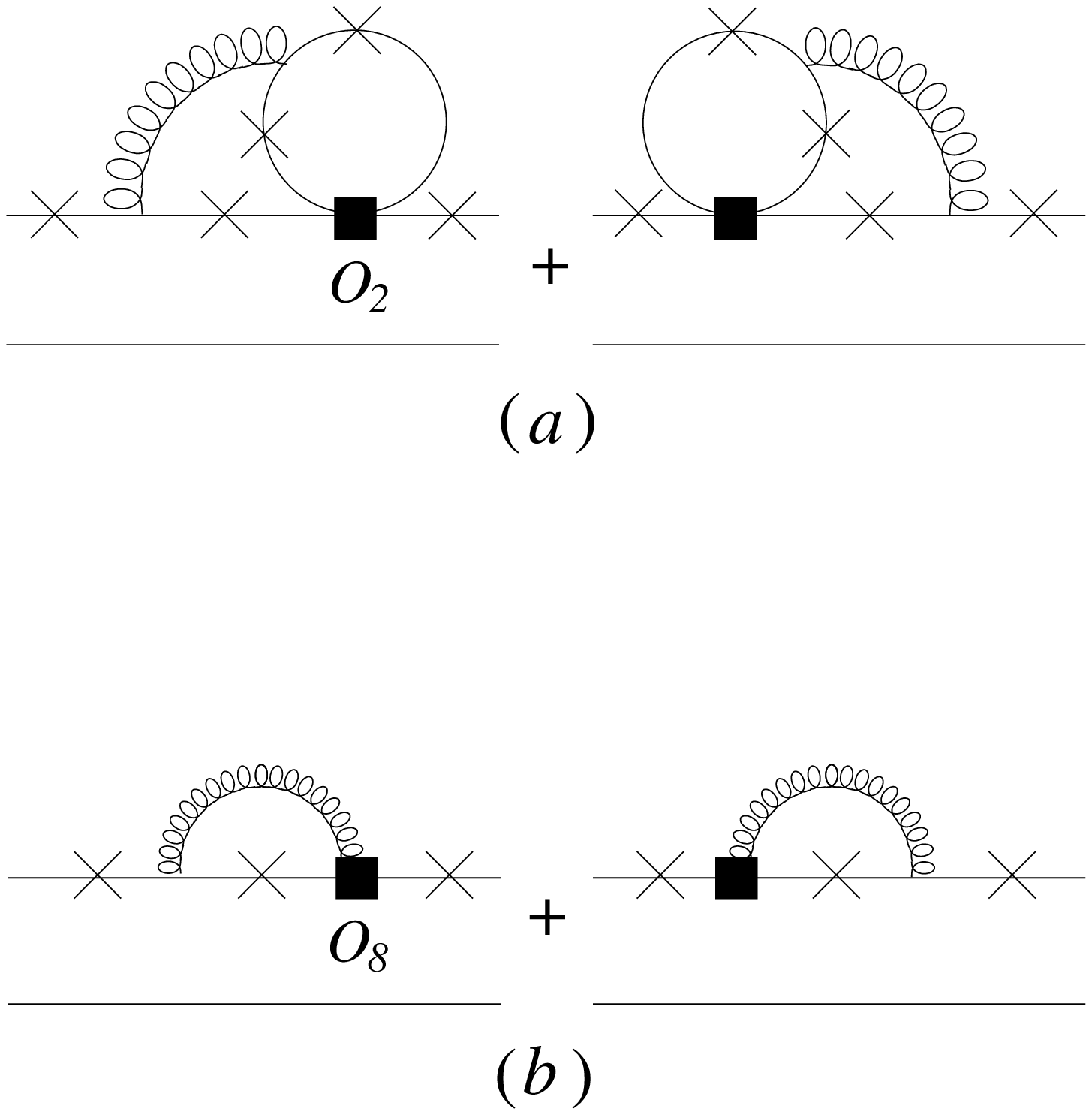}
\end{center}
\caption{}
\label{VC}
\end{figure}

 
The vertex corrections are directly proportional to the form factor $F^A_+$.
They are given by (Fig.\ \ref{VC}) \cite{Hurth,Chetyrkin}
\begin{eqnarray}
\langle O_2\rangle_{VC}&=&
 \frac{\alpha_s}{4\pi}\langle O_7\rangle\Bigg(
  \frac{416}{81}\ln\frac{m_b}{\mu}+r_2\Bigg)~,\\
\langle O_8\rangle_{VC}&=&\frac{\alpha_s}{4\pi}\langle O_7\rangle\Bigg[
   -\frac{32}{9}\ln\frac{m_b}{\mu}+\frac{4}{27}(33-2\pi^2+6i\pi)\Bigg]~,
\end{eqnarray}
where
\begin{eqnarray}
r_2&=&\frac{2}{243}\Big\{-833+144\pi^2 z^{3/2}\nonumber\\
 &&+\Big[1728-180\pi^2-1296\zeta(3)
 +(1296-324\pi^2)L+108L^2+36L^3\Big]z\nonumber\\
 &&+\Big[648+72\pi^2+(432-216\pi^2)L+36L^3\Big]z^2
   +\Big[-54-84\pi^2+1092L-756L^2\Big]z^3\Bigg\}\nonumber\\
 &&+i\frac{16\pi}{81}\Bigg\{-5+\Big[45-3\pi^2+9L+9L^2\Big]z
   +\Big[-3\pi^2+9L^2\Big]z^2+\Big[28-12L\Big]z^3\Bigg\}~,
\end{eqnarray}
with $z\equiv m_c^2/m_b^2$, $L\equiv \ln z$, and $\zeta(x)$ being the Liemann
$\zeta$-function.
\par
Hard spectator corrections are well described by the convolution
between the hard kernel $T_i(\xi,u)$ and the light-cone distribution amplitudes 
of the involved mesons, $\Phi_B(\xi)$ and $\Phi_A(u)$, in the heavy quark limit;
\begin{equation}
\langle O_i\rangle_{HS}=\int_0^1 d\xi du \Phi_B(\xi)T_i(\xi,u)\Phi_A(u)~.
\end{equation}
\par
The light-cone distribution amplitudes are defined by
\begin{mathletters}
\label{DA}
\begin{eqnarray}
\langle 0| b(0){\bar q}'(z)|B(p)\rangle&=&\frac{if_B}{4}
 ({p\hspace{-2.0mm}/}+m_B)\gamma_5\int_0^1 d\xi~ e^{-i\xi p_+z_-}\Big[
 \Phi_{B1}(\xi)+{\nbar\hspace{-2.1mm}/}\Phi_{B2}(\xi)\Big]~,\\
\langle A(p',\epsilon)|q(z)\qbar(0)|0\rangle&=&\frac{f_A^\perp}{4}\gamma_5
 \sigma^{\mu\nu}\epsilon_\mu p'_\nu\int_0^1 du~ e^{i\ubar p'\cdot z}
 \Phi_A^\perp(u)~,~~~~~(\ubar\equiv 1-u)
\label{ADA}
\end{eqnarray}
where $\nbar^\mu=(1,0,0,-1)$ is parallel to the outgoing meson.
\end{mathletters}
To calculate the hard spectator contributions, following kinematics for 
Fig.\ \ref{O7NLO} is adopted:
\begin{eqnarray}
p_b^\mu&=&m_b v^\mu~,\nonumber\\
l^\mu&=&\frac{l_+}{2}n^\mu+l^\mu_\perp+\frac{l_-}{2}\nbar^\mu~,\nonumber\\
q^\mu&=&\omega n^\mu~~~(\omega\simeq m_B/2)~,\nonumber\\
k_1^\mu&\simeq&uE\nbar^\mu+k_\perp^\mu+\calO(k_\perp^2)~,\nonumber\\
k_2^\mu&\simeq&\ubar E\nbar^\mu-k_\perp^\mu+\calO(k_\perp^2)~~~(E\simeq m_B/2)~,
\end{eqnarray}
where $n^\mu=(1,0,0,1)$ and $u$ is the relative energy fraction.
\par
\begin{figure}
\begin{center}
\epsfig{file=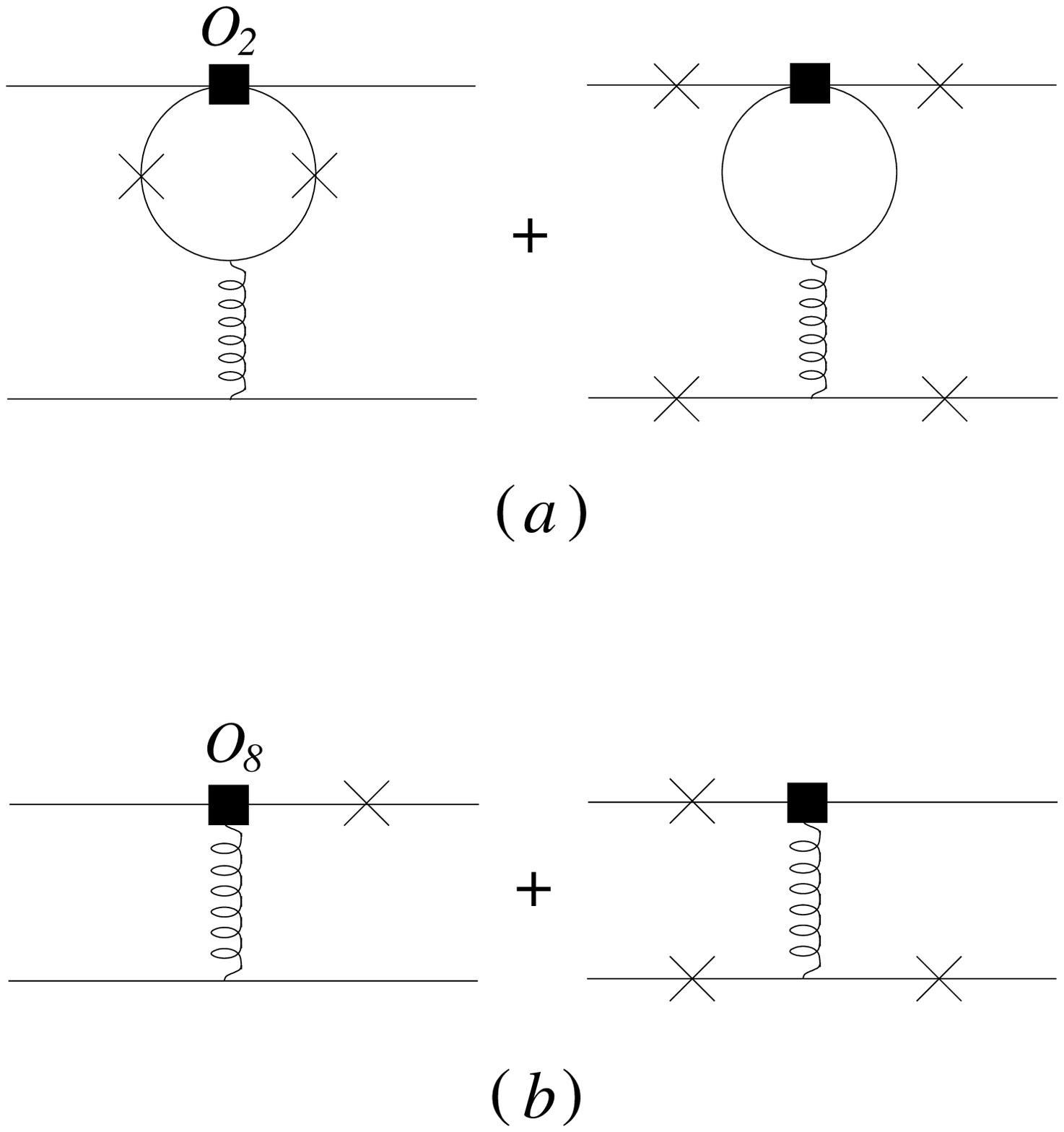}
\end{center}
\caption{}
\label{HS}
\end{figure}

Direct calculation of each diagram in Fig.\ \ref{HS} plugged with 
Eq.\ (\ref{DA}) yields
\begin{eqnarray}
\label{O28HS}
\langle O_2\rangle_{HS}&=&
 \frac{\langle O_7\rangle_A}{F^A_+(0)}\frac{4\pi\alpha_s C_F}{N_c}
 \frac{f_B f_A^\perp}{m_b m_B}\Bigg[
  \frac{1}{12}\langle u^{-1}\Delta F_1(z_1^{(c)})\rangle_\perp
 +\frac{3}{16}Q_{sp}\langle\ubar^{-1}\Delta F_1(z_0^{(c)})\rangle_\perp
\nonumber\\
&&
  -\frac{1}{12}\langle\xi^{-1}\rangle_1
   \langle\ubar^{-1}\Delta i_5(z_0^{(c)},z_1^{(c)},0)\rangle_\perp
  -\frac{1}{3}
   \langle\ubar^{-1}\Delta i_{25}(z_0^{(c)},z_1^{(c)},0)\rangle_\perp\Bigg]~,\\
\langle O_8\rangle_{HS}&=&
 \frac{\langle O_7\rangle_A}{F^A_+(0)}\frac{4\pi\alpha_s C_F}{N_c}
 \frac{f_B f_A^\perp}{m_B^2}\Bigg[
 \frac{1}{12}\langle u^{-1}\rangle_\perp\langle\xi^{-1}\rangle_1
 +\frac{Q_{sp}}{8}(\langle\ubar^{-1}\rangle_\perp
   +2\langle\ubar^{-2}\rangle_\perp)\Bigg]
~.
\end{eqnarray}
Here $N_c$ is the number of color with $C_F=\frac{N_c^2-1}{2N_c}$, and $Q_{sp}$
is the electric charge of the spectator quark.
The expectation values over the distribution amplitudes are defined by
\begin{mathletters}
\begin{eqnarray}
\langle f(u)\rangle_\perp&\equiv&\int_0^1 du~f(u)\Phi_A^\perp(u)~,\\
\langle\xi^N\rangle_1&\equiv&\int_0^1 d\xi~\xi^N \Phi_{B1}(\xi)~.
\end{eqnarray}
\end{mathletters}
Relevant functions $\Delta F_1$, $\Delta i_5$, and $\Delta i_{25}$ as well
as the arguments $z_{0,1}^{(f)}$ are given in \cite{Ali,Simma}.

\section{Branching ratios for $B\to K_1\gamma$}

The branching ratio of $B\to K_1\gamma$ is simply given by
\begin{equation}
\calB(B\to K_1\gamma)=\tau_B\frac{G_F^2\alpha m_b^2 m_B^3}{32\pi^4}\Bigg(
 1-\frac{m_A^2}{m_B^2}\Bigg)^3|F^A_+(0)|^2|V_{tb}V_{ts}^*|^2 
 |C_7^{\rm eff}(\mu_b)+A_{VC}+A_{HS}|^2~.
\end{equation}
At the heavy quark limit,
\begin{eqnarray}
A_{VC}&=&\frac{\alpha_s(\mu_b)}{4\pi}\Bigg\{C_8^{\rm eff}(\mu_b)\Bigg[
 -\frac{32}{9}\ln\frac{m_b}{\mu_b}+\frac{4}{27}(33-2\pi^2+6i\pi)\Bigg]
 +C_2(\mu_b)\Bigg[\frac{416}{81}\ln\frac{m_b}{\mu_b}+r_2\Bigg]\Bigg\}~,
\nonumber\\
A_{HS}&=&\frac{4\pi\alpha_s(\mu_H)C_F}{N_c}
 \frac{f_B f_A^\perp}{\lambda_B m_B F^A_+(0)}\Bigg\{
 C_8^{\rm eff}(\mu_H)\frac{1}{12}\langle u^{-1}\rangle_\perp
 -C_2(\mu_H)\frac{1}{12}\left\langle
 \frac{\Delta i_5(z_0^{(c)},0,0)}{\ubar}\right\rangle_\perp\Bigg\}~,
\end{eqnarray}
where the negative moment of $\Phi_{B1}$ is parameterized by 
$\lambda_B\sim\calO(\Lambda_{\rm QCD})$ as
\begin{equation}
\int_0^1d\xi~\frac{\Phi_{B1}(\xi)}{\xi}\equiv\frac{m_B}{\lambda_B}~.
\end{equation}
The renormalization scale is fixed at $\mu=\mu_b=\calO(m_b)$ for the vertex
corrections while for the hard spectator interactions, 
$\mu=\mu_H\sim\sqrt{\Lambda_{\rm QCD}m_b}$.
In the following analysis, we set $\mu_b=m_b$ and $\mu_H=\sqrt{\Lambda_H m_b}$
where $\Lambda_H=0.5$ GeV.
\par
The scale dependence of $\langle O_7\rangle$ is absorbed into the product of
$b$-quark mass and the form factor; \cite{Bosch}
\begin{equation}
(m_b\cdot F^A_+)[\mu]=(m_b\cdot F^A_+)[m_b]\Bigg(1+\frac{\alpha_s(\mu)}{4\pi}
\frac{32}{3}\ln\frac{m_b}{\mu}\Bigg)~.
\end{equation}
Other input values are summarized in Table \ref{inputs}.
Contrary to the $B\to K^*\gamma$, there are few reliable values for $F^A_+(0)$
and $f_A^\perp$ both in theory and experiment in the literature.
We adopt the results from the light-cone sum rules by Safir \cite{Safir}, 
whose values are listed in Table \ref{LCSR}.
In Table \ref{componential}, each contributions to the decay amplitudes is 
listed from the central values of Tables \ref{inputs} and \ref{LCSR}.
Note that the NLO corrections contribute positively, except $C_7^{\rm eff(1)}$.
Reference scale for the present analysis is 
\begin{equation}
(\mu_b,\mu_H)=(m_b(m_b), \sqrt{\Lambda_H m_b(m_b)})
=(4.2~{\rm GeV}, 1.45~{\rm GeV})~.
\end{equation}
As a comparison, results for another scale $(\mu_b,\mu_H)=(m_{b,PS}, (m_b(m_b))$
are also given in Table \ref{componential}, where $m_{b,PS}=4.6$ GeV is the 
so-called potential-subtracted mass \cite{PS}.
It should be emphasized that in Table \ref{componential}, $C_7^{\rm eff}$ and
$A_{VC}$ are process independent, and encodes QCD effects only.
On the other hand, $A_{HS}$ contains the key information of the outgoing meson.
Although $F^A_+(0)$ in $A_{HS}$ is canceled, non-perturbative properties of 
daughter meson still remain in $f_A^\perp$ and $\langle\cdots\rangle_\perp$.
When averaging over $\Phi_A^\perp (u)$, process dependence is encapsulated in
the coefficients of the Gegenbauer expansion, which vanish at $\mu\to\infty$.
We simply neglect the expansion here, retaining $\Phi_A^\perp$ as its asymptotic
form
\begin{equation}
\Phi_A^\perp (u)\approx\Phi_A^{\perp (as)}(u)=6u\ubar~.
\end{equation}
Keeping the hadronic parameters specifically, we have
\begin{eqnarray}
\calB(B^0\to K_1^0\gamma)&=&0.003\times\Bigg(1-\frac{m^2}{m_B^2}\Bigg)^3\times
 \left|
 F^A_+(0)(-0.385-i0.014)\right.\nonumber\\
&& \left.+(f_A^\perp/{\rm GeV})(-0.024-i0.022)\right|^2~.
\label{master}
\end{eqnarray}
Final results for the decay amplitudes and the branching ratios are listed
in Table \ref{scale}.
Uncertainties in the branching ratios are from those in the form factor.
For the charged modes, one has only to multiply the life-time ratio 
$\tau_{B^\pm}/\tau_{B^0}$ to the above equation.
\par
In Eq.\ (\ref{master}), the coefficient of $F^A_+(0)$ is 
$C_7^{\rm eff}(\mu_b)+A_{VC}(\mu_b)$, while that of $f_A^\perp$ is 
$A_{HS}(\mu_H)\times F^A_+(0)/f_A^\perp$.
Since the presence of $\gamma_5$ in Eq.\ (\ref{ADA}) does not change the trace
calculation for getting Eq.\ (\ref{O28HS}) and the form of $\Phi_A^{\perp(as)}$
is universal, the numerics in Eq.\ (\ref{master})
are common to both $B\to K_V\gamma$ and $B\to K_A\gamma$, irrespective of the 
species of $K_V$ or $K_A$.
This is quite an interesting point considering the fact that the measurements
for $B\to K_A\gamma$ are near at hand.
Most of all, the mass hierarchy of $m_{K^*}<1~{\rm GeV}< m_{K_1}$ might impose
some doubts about the common framework for both $K^*$ and $K_1$.
Actually, the scale 1 GeV is very delicate because the chiral symmetry is 
broken around it.
Recall that in calculating the hard spectator interactions it is assumed that
the axial Kaon is nearly massless.
Although the assumption is acceptable for $m_{K_1}\ll m_B$, it is also possible
that nonzero mass effects are sizable.
So far, there is no systematics to deal with it.
The compatibility of Eq.\ (\ref{master}) with experimental observations for both
$B\to K^*\gamma$ and $B\to K_1 \gamma$ will cast some clues to this issue.
In the kinematically opposite limit where $K_1$ is very heavy, 
Ref.\ \cite{Ohl,Veseli} predicted branching ratios of higher Kaon resonances.
Their results as well as those from other methods are listed in Table 
\ref{comparison} for a comparison.
In the heavy quark scheme, hard spectator interaction is inconceivable since 
almost all the momentum of initial heavy meson is transfered to the final one.
Typical scale of interaction with the spectator is $\sim\Lambda_{\rm QCD}$ where
the perturbative approach breaks down.
Thus checking the validity of hard spectator contribution plays an important
role in determining which approach is more reliable.
\par
The biggest uncertainty in theoretical prediction lies in calculation of the 
form factor $F^A_+$.
QCD sum rule is among the most reliable.
But recent analysis on $B\to K^*\gamma$ reveals that LCSR
results for the relevant form factor lead to a very large branching ratio
compared to the measured one \cite{Ali}.
Unfortunately, there is no way to explain the discrepancy up to now.
The will-be-extracted values of $F^A_+$ from the experiments, therefore, 
provide much interest to see whether the LCSR predicts larger form factors 
again.
\par
Another issue of $B\to K_1\gamma$ is mixing.
If experiments measure very different values of $\calB(B\to K_1(1270)\gamma)$
and $\calB(B\to K_1(1400)\gamma)$, then the maximal mixing of $K_{1A}$ and
$K_{1B}$, which correspond to $^3P_1$ and $^1P_1$ quark model states
respectively, is more favored \cite{HYCheng}.
One can be about 40 times larger than the other.
\par 
Present analysis is done at the heavy quark limit,
at NLO of $\alpha_s$, and at the leading twist of the distribution amplitudes 
for the involved mesons.
At the heavy quark limit, only the terms proportional to 
$\langle\xi^{-1}\rangle_1\sim\calO(1/\Lambda_{\rm QCD})$ survive. 
And the NLO $\alpha_s$ effects are,
\begin{equation}
\frac{|C_7^{\rm eff(0)}|^2}{|C_7^{\rm eff}+A_{VC}+A_{HS}|^2}\approx 62\%~,
\end{equation}
for both $K_1(1270)$ and $K_1(1400)$ at 
$(\mu_b,\mu_H)=(4.2~{\rm GeV}, 1.45~{\rm GeV})$.
Higher twist effects are nontrivial and process dependent in general.
For $B\to K^*\gamma$, the non-asymptotic correction of $K^*$ at higher twist
through the Gegenbauer moments to the operator $O_8$ amounts to $\sim -20\%$ 
\cite{Ali}.
Similar effects are expected in $K_1$.

\section{Conclusions}

Radiative $B$ decays to the Kaon resonances provide a rich laboratory to test
the standard model and probe new physics.
$B\to K^*\gamma$ is a well established process, and
Belle and BaBar are now measuring the decay modes of higher resonances for the
first time.
In a theoretical side, deeper understandings have been accomplished for a
decade.
For example, relevant Wilson coefficients are known up to the three-loop level.
The idea of the QCD factorization reduces model or process dependences.
And various versions of effective theories of QCD such as HQET or SCET have
simplified the analysis dramatically.
\par
In this paper, radiative $B$ decays to the axial Kaons are examined at NLO of
$\calO(\alpha_s)$.
This was already done for $K^*$ a few years ago, and many aspects are common.
Especially, they share the same perturbative QCD part and only the weak form
factor as well as some static properties of the final $K_{\rm res}$ discern
the specific process, at the leading twist and heavy quark limit.
\par
On the other hand, the largest uncertainty of theory is the form factor for
which we used the LCSR calculations.
Since the results of LCSR for $B\to K^*$ form factor turn out to be quite
large compared to the experiments, the reliability is rather low.
A clear explanation of the discrepancy will remain a good challenge.
In this respect, near future measurements for $B\to K_1\gamma$ and extraction
of the form factor are quite exciting.
They also check the possible mixing between $^3 P_1$ and $^1P_1$ states to 
form physical $K_1(1270)$ and $K_1(1400)$.

\begin{center}
{\large\bf Acknowledgements}\\[10mm]
\end{center}

The author thanks Heyoung Yang and Mikihiko Nakao for their reading the 
manuscript and giving comments.
This work was supported by the BK21 Program of the Korean Ministry of Education.


\newpage

\begin{center}{\large\bf FIGURE CAPTIONS}\end{center}

\noindent
Fig.\ 1
\\
Leading order contribution by operator $O_7$.
\vskip .3cm

\noindent
Fig.\ 2
\\
NLO corrections to $O_7$.
These diagrams are absorbed into the weak form factor $F^A_+$.
\vskip .3cm

\noindent
Fig.\ 3
\\
Vertex corrections to the operators (a) $O_2$ and (b) $O_8$.
Crosses denote the possible attachment of the emitted photon.
\vskip .3cm

\noindent
Fig.\ 4
\\
Hard spectator interactions to (a) $O_2$ and (b) $O_8$.
First diagrams are leading contributions at the heavy quark limit.
\vskip .3cm

\par

\pagebreak


\begin{figure}
\vskip 2cm
\begin{center}
\epsfig{file=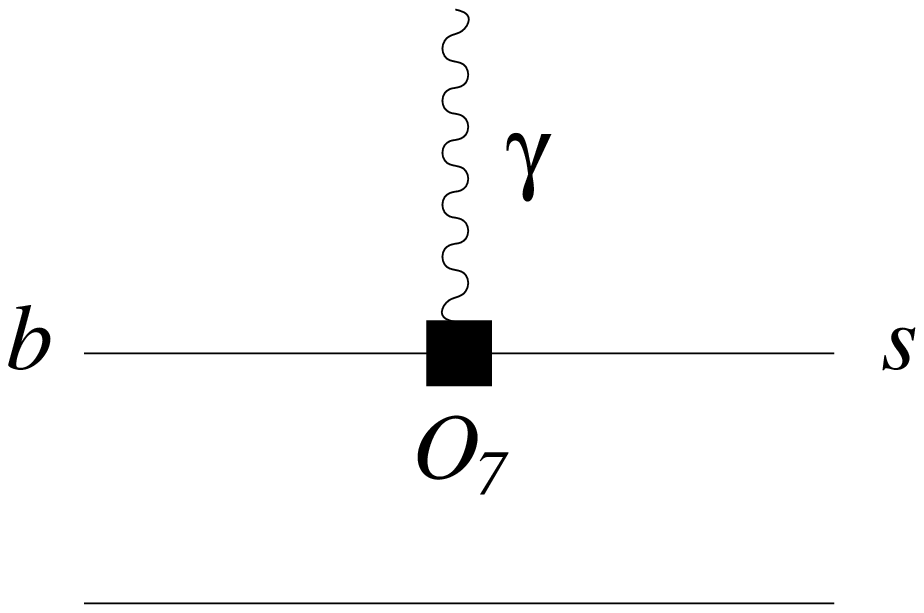}
\end{center}
\caption{}
\label{O7}
\end{figure}



\pagebreak


\begin{table}
\caption{Summary of input values}
\begin{tabular}{cc}
$|V_{tb}V_{ts}^*|$ & $0.0396\pm0.0020$ \cite{Ali} \\
$\tau_{B^+}$ & $(1.674\pm0.018)$ ps \\
$\tau_{B^0}$ & $(1.542\pm0.016)$ ps \\
$m_B$ & 5.28 GeV \\
$f_B$ & 0.18 GeV \\
$\lambda_B$ & $(0.35\pm0.15)$ GeV \\
$m_b(m_b)$ & 4.2 GeV \\
$m_c(m_b)$ & $(1.3\pm0.2)$ GeV \\
\end{tabular}
\label{inputs}
\end{table}


\begin{table}
\caption{$F^A_+(0)$ and $f_A$ from light-cone sum rules}
\begin{tabular}{c|cc}
Axial $K_1$ & $K_1(1270)$ & $K_1(1400)$ \\\hline
$m_A$ & 1.273 GeV & 1.402 GeV \\
$f_A$ & 0.122 GeV & 0.091 GeV \\
$F^A_+(0)$ & $0.14\pm0.03$ & $0.098\pm0.02$\\
\end{tabular}
\label{LCSR}
\end{table}


\begin{table}
\caption{Componential contributions to the decay amplitude}
\begin{tabular}{c|cc}
$\mu_b$ & $m_b(m_b)=4.2$ GeV & $m_{b,PS}=4.6$ GeV \\\hline
$C_7^{\rm eff(0)}(\mu_b)$ & $-0.321$ & $-0.316$ \\
$C_7^{\rm eff(1)}(\mu_b)$ & $0.602$ & $0.522$ \\
$C_7^{\rm eff}(\mu_b)$ & $-0.310$ & $-0.307$ \\
$A_{VC}(\mu_b)$ & $-0.075-i0.014$ & $-0.082-i0.013$ \\\hline\hline
$\mu_H$ & $\sqrt{\Lambda_H m_b(m_b)}=1.45$ GeV & $m_b(m_b)=4.2$ GeV \\\hline
$A_{HS}^{K1(1270)}(\mu_H)$ & $-0.021-i0.019$ & $-0.013-i0.013$ \\
$A_{HS}^{K1(1400)}(\mu_H)$ & $-0.022-i0.020$ & $-0.014-i0.013$ \\
\end{tabular}
\label{componential}
\end{table}


\begin{table}
\caption{Decay amplitudes and branching ratios for different scales}
\begin{tabular}{c|cccc}
$(\mu_b,\mu_H)$ (GeV)&$(4.2,1.45)$&$(4.2,4.2)$&$(4.6,1.45)$&$(4.6,4.2)$ 
\\\hline\hline
$(C_7^{\rm eff}+A_{VC}+A_{HS})_{K1(1270)}$ & 
 $-0.406-i0.033$ & $-0.399-i0.027$ & $-0.410-i0.033$ & $-0.402-i0.026$ \\
$\calB(B^0\to K_1^0(1270)\gamma)\times 10^5$ & 
 $0.828\pm0.335$ & $0.795\pm0.329$ & $0.814\pm0.341$ & $0.782\pm0.335$ \\\hline
$(C_7^{\rm eff}+A_{VC}+A_{HS})_{K1(1400)}$ & 
 $-0.408-i0.034$ & $-0.400-i0.027$ & $-0.412-i0.034$ & $-0.403-i0.027$ \\
$\calB(B^0\to K_1^0(1400)\gamma)\times 10^5$ & 
 $0.393\pm 0.151$ & $0.376\pm0.148$ & $0.386\pm0.154$ & $0.370\pm0.150$
\end{tabular}
\label{scale}
\end{table}


\begin{table}
\caption{Comparison with other results, in units of $10^{-5}$.}
\begin{tabular}{c||cc}
Branching Ratio 
& $\calB(B\to K_1(1270)\gamma)$
& $\calB(B\to K_1(1400)\gamma)$\\\hline
JPL & $0.828$ & $0.393$ \\
Ref.\ \cite{HYCheng}& $0.02\sim0.84$ & $0.003\sim0.80$ \\
Ref.\ \cite{Safir}& $0.493$ & $0.241$ \\
Ref.\ \cite{Ebert}& $0.45$ & $0.78$  \\
Ref.\ \cite{Veseli}& $1.20$ & $0.58$  \\
Ref.\ \cite{Atwood}& $0.3\sim1.4$ & $0.1\sim0.6$  \\
Ref.\ \cite{Ohl}& $1.8\sim4.0$ & $2.4\sim5.2$  \\
Ref.\ \cite{Altomari}& $1.1$ & $0.7$  \\
\end{tabular}
\label{comparison}
\end{table}


\begin{thebibliography}{99}

\bibitem{Belle}
Belle Collaboration, K.\ Abe \etal, BELLE-CONF-0319 (2003).
\bibitem{BaBar}
BaBar Collaboration, B.\ Aubert \etal, \prl{88}{2002}{101805}.
\bibitem{CLEO}
CLEO Collaboration, T.E.\ Coan \etal, \prl{84}{2000}{5283}.
\bibitem{Soares}
J.M.\ Soares, \npb{367}{1991}{575}; \prd{49}{1994}{283}.
\bibitem{Greub}
C.\ Greub, H.\ Simma, and D.\ Wyler, \npb{434}{1995}{39}; 
Erratum \ibid{444}{1995}{447}.
\bibitem{Hurth}
C.\ Greub, T.\ Hurth, and D.\ Wyler, \prd{54}{1996}{3350}.
\bibitem{Adel}
K.\ Adel and Y.\ Yao, \prd{49}{1994}{4945}.
\bibitem{Chetyrkin}
K.\ Chetyrkin, M.\ Misiak, and M.\ M\"unz, \plb{400}{1997}{206}.
\bibitem{BBNS}
M.\ Beneke, G.\ Buchalla, M.\ Neubert, and C.T.\ Sachrajda, 
\npb{591}{2000}{313}.
\bibitem{Feldmann}
M.\ Beneke and T.\ Feldmann, \npb{592}{2001}{3}.
\bibitem{Siedel}
M.\ Beneke, T.\ Feldmann, and D.\ Siedel, \npb{612}{2001}{25}.
\bibitem{Bosch}
S.W.\ Bosch and G.\ Buchalla, \npb{621}{2002}{459}.
\bibitem{Ali}
A.\ Ali and A.Ya.\ Parkhomenko, \epjc{23}{2002}{89}.
\bibitem{Chay}
J.\ Chay and C.\ Kim, \prd{68}{2003}{034013}.
\bibitem{Gronau}
M.\ Gronau, Y.\ Grossman, D.\ Pirjol, and A.\ Ryd, \prl{88}{2002}{051802};
M.\ Gronau and D.\ Pirjol, \prd{66}{2002}{054008}.
\bibitem{jplee}
J.-P.\ Lee, \prd{69}{2004}{014017}; 
{\it Proceeding of the 2nd ICFP 03, Seoul, Korea} [hep-ph/0312010].
\bibitem{Belle2}
Belle Collaboration, S.\ Nishida \etal, \prl{89}{2002}{231801}.
\bibitem{BaBar2}
BaBar Collaboration, A.\ Aubert \etal, [hep-ex/0308021].
\bibitem{Ohl}
A.\ Ali, T.\ Ohl, and T.\ Mannel, \plb{298}{1993}{195}.
\bibitem{Veseli}
S.\ Veseli and M.G.\ Olsson, \plb{367}{1996}{309}.
\bibitem{Altomari}
T.\ Altomari, \prd{37}{1998}{677}
\bibitem{Atwood}
D.\ Atwood and A.\ Soni, \zpc{64}{1994}{241}.
\bibitem{Ebert}
D.\ Ebert, R.N.\ Faustov, V.O.\ Galkin, and H.\ Toki, \prd{64}{2001}{054001}.
\bibitem{HYCheng}
H.-Y.\ Cheng and C.-K.\ Chua, [hep-ph/0401141].
\bibitem{Safir}
A.S.\ Safir, Eur. Phys. J. direct C {\bf 3}, 15 (2001).
\bibitem{Simma}
H.\ Simma and D.\ Wyler, \npb{344}{1990}{283}.
\bibitem{PS}
M.\ Beneke, \plb{434}{1998}{115};
M.\ Beneke, A.\ Signer, \plb{471}{1999}{233}.

\end{thebibliography}
\end{document}